\documentstyle[12pt]{article}
\newcommand{\half}{\frac 1 2 }

\newcommand{\ie}{{\em i.e.} }
\newcommand{\be}{\begin{eqnarray}}
\newcommand{\ee}{\end{eqnarray}}

\catcode`\@=11
\newcommand{\ccite}[1] {\@ifundefined{b@#1}{\bf ?}{\@nameuse{b@#1}}}
\catcode`\@=12

\begin{document}
\vspace*{1cm}
\centerline{\Large\bf On the anyon description of the Laughlin hole
states}
\vspace* {-45 mm}
\begin{flushright} Oslo SHS-96-6 \\ June 1996 \end{flushright}
\vskip 55mm
\centerline{\bf Heidi Kj{\o}nsberg$^{\dagger}$$^{\ddagger}$and Jon Magne
Leinaas$^{\star}$  } \medskip \centerline{Centre for Advanced
Study}
\centerline{at the Norwegian Academy of Science and Letters,}
\centerline{P.O. Box 7606 Skillebekk, N-0205 Oslo, Norway} 
\centerline{and}
\centerline{Department of Physics, University of Oslo} 
\centerline{P.O. Box
1048 Blindern, N-0316 Oslo, Norway}
\vskip 15mm
\def\fk{\mbox{ $f_K$} }
\centerline{\bf ABSTRACT}
\vspace{.5cm}
We examine the anyon representation of the Laughlin quasi-holes, in 
particular
the one-dimensional, algebraic aspects of the representation. For the 
cases of
one and two quasi-holes an explicit mapping to anyon systems is given, and
the connection between the hole-states and coherent states of the 
fundamental
algebras of observables is examined. The quasi-electron case is 
discussed more
briefly, and some remaining questions are pointed out.
\medskip
\vskip 3mm

\vfil
\noindent
$^{\dagger}$e-mail: heidi.kjonsberg@fys.uio.no\\ $^{\star}$e-mail:
j.m.leinaas@fys.uio.no\\
$^{\ddagger}$Supported by The Norwegian Research Council.

\eject
\section{Introduction}

In Laughlin's theory of the fractional quantum Hall effect 
\cite{kn:laugh1}
the quasi-holes and quasi-electrons satisfy fractional statistics
\cite{kn:leinmyr}. They are
anyons in a strong magnetic field. The fractional statistics property of
these excitations was first suggested by Haldane \cite{kn:hald} and 
Halperin
\cite{kn:halp} in the construction of the hierarchy of Hall states. It was
demonstrated more directly by Arovas, Schrieffer and Wilczek, who 
considered a pair of quasi-holes encircling
each other and interpreted the two-particle contribution to the Berry 
phase
as a statistics phase \cite{kn:arovas}. Laughlin later suggested an 
explicit
particle representation of the quasi-holes and quasi-electrons where these
have anyonic properties \cite{kn:Lau,kn:laugh2}.

Due to the strong magnetic field, the quasi-holes and quasi-electrons
effectively behave like particles in one dimension. The relevant
coordinates are the particle coordinates projected to the lowest Landau
level, which are the guiding center coordinates of the particles. These do
not commute, and the physical plane gets the character of a 
two-dimensional
phase space. In fact the quasi-holes and quasi-electrons are more strictly
one-dimensional than the electrons of the system, since excitations of
these
to higher Landau levels are missing. This suggests that the hole and
particle excitations may be characterized by {\em one-dimensional}
fractional statistics as an alternative to the two-dimensional anyon
statistics referred to above.

An algebraic approach to fractional statistics in one dimension was
introduced in ref.\cite{kn:joja}. For a single particle the Hilbert space
defines an irreducible representation of a Heisenberg-Weyl (HW) algebra of
observables, the algebra defined by the coordinate and momentum. For two
{\em identical} particles the fundamental algebra of observables does not
consist of two independent HW algebras. This is due to the symmetrization
postulate for the observables. Instead the algebra may be viewed as
consisting of a HW algebra for the center-of-mass coordinates and an
$su(1,1)$ algebra for the relative coordinates. The latter is quadratic in
the coordinates. The irreducible representations of the
$su(1,1)$ algebra are characterized by a real parameter which is 
interpreted
as the one-dimensional statistics parameter.

In ref.\cite{kn:hansleinmyr} this one-dimensional algebraic approach was
applied to a two-anyon system in the lowest Landau level. It was shown
explicitly how to construct the raising and lowering operators of the
$su(1,1)$ algebra and a linear relation was found between the
one-dimensional statistics parameter and the (two-dimensional) anyon
parameter. For bosons and fermions the generators of the $su(1,1)$ were
quadratic in anyon coordinates and derivatives, but in the general
case 
the
expressions were more complicated.

The purpose of this paper is to examine the anyon representation of the
quasi-holes, in particular the one-dimensional aspects, somewhat
closer. 
We
restrict the discussion mainly to one and two quasi-holes, although with
some comments also about the quasi-electron case. We follow the ideas of
Laughlin and construct an explicit anyon representation of the 
quasi-holes.
We do it in a rather detailed way and focus special attention on the
algebraic properties of the one-dimensional system. A particular 
question is
whether the quasi-holes can be viewed as coherent states of the 
fundamental
(one-dimensional) algebra of observables. The discussion is closely 
related
to that of ref.\cite{kn:hansleinmyr}, but now with application to
quasi-holes of the quantum Hall system rather than to a general anyon 
system.

\section{One particle --- one quasi-hole.} In this section we consider the
case of a single quasi-hole and its relation to the single particle in a
strong magnetic field. We first have a brief discussion of the one 
particle
case and in the next subsection consider the explicit mapping between the
quasi-hole of the Laughlin theory and the single particle system.
\subsection{The one particle case.}

For a single, planar particle in a magnetic field, the Hamiltonian and
conserved angular momentum are
\begin{equation} H= \frac{m}{2}\; {\bf v}^{2}, \hspace{.5cm} L=m({\bf r}
\times {\bf v})_{z} + \frac{1}{2}m\omega{\bf r}^{2}; \hspace{.5cm} m{\bf
v}={\bf p} - \frac{q}{c}{\bf A}. \label{eq:start}
\end{equation} The charge \(q\) is assumed to be positive, and 
\({\bf B} =
B{\bf e}_{z}\) with \(B>0\), so the cyclotron frequency \(\omega =
\frac{qB}{mc}\) is positive and the magnetic length is \(l_{B} =
(\frac{\hbar c}{qB})^{1/2}\). In terms of the guiding center coordinate
\({\bf R}\) and the relative coordinate
\({\bf r}_{rel}\) 
(using the notation \( ^{\star}{\bf v} = (-v_{y},v_{x})\) )
\begin{equation} {\bf R} = {\bf r} -
\frac{1}{\omega}\hspace{.1cm}^{\star}{\bf v} , \hspace{1cm} 
{\bf r}_{rel} = -
\frac{1}{\omega}\hspace{.1cm}^{\star}{\bf v}, \label{eq:guirel}
\end{equation} a convenient set of variables is defined by the 
dimensionless
operators \begin{equation} a = \frac{1}{{\sqrt 2}l_{B}}(X-iY) , 
\hspace{1cm}
b = \frac{1}{{\sqrt 2}l_{B}}(x_{rel} + i y_{rel} ), \label{eq:comp}
\end{equation} and their adjoints. The Hamiltonian and angular
momentum 
are
then given by the well known gauge independent expressions 
\begin{equation}
H= \hbar \omega (b^{\dagger} b + \frac{1}{2}) , \hspace{1cm} L=
\hbar(a^{\dagger}a - b^{\dagger}b).
\end{equation} The only non-vanishing commutation relations are
\begin{equation} [b,b^{\dagger}] = [a, a^{\dagger}] = 1,
\end{equation} and the system thus consists of two uncoupled 
one-dimensional
harmonic oscillators. The harmonic oscillator states 
\(\mid\!l,n \rangle \),
\(n,l =0,1,_{\cdots}\), correspond to energies \(E_{n} = \hbar\omega
(n+\frac{1}{2}) \) and angular momentum eigenvalues \(L = (l-n)\) 
(measured in
units of \(\hbar\)), and the variable $n$ then specifies a given 
Landau level.

In each specific energy level (Landau level) the fast relative 
coordinates
are frozen out, and the resulting system has one-dimensional
dynamics. 
The
corresponding state space yields (in the unbounded case) a 
representation of
the Heisenberg-Weyl algebra \(\{a,a^{\dagger},1\}\). The coherent 
states of
this algebra are defined by 
\begin{equation} a\mid \! \alpha^{*},n \rangle =
\alpha^{*} \mid \! \alpha^{*},n \rangle.
\end{equation} The state can be written as
\begin{eqnarray}
\mid \! \alpha^{*},n \rangle &=& e^{-\frac{1}{2} \alpha \alpha^{*} } 
\sum_{l
=0}^{\infty} \frac{(\alpha^{*})^{l}}{{\sqrt l!}} \mid \! l,n\rangle
\nonumber \\
\noalign{\medskip} &=& e^{-\frac{1}{2}\alpha\alpha^{*} + \alpha^{*}
a^{\dagger}} \mid\!0,n\rangle. \label{eq:scser}
\end{eqnarray} One can show that for each specific Landau level \(n\) the
uncertainty in position generally must obey \(\Delta x \Delta y \geq
l_{B}^{2} (n+1)\). The coherent states satisfy the equality sign in this
relation, and in this sense they are maximally localized in the physical
plane.

Introducing the complex particle coordinate 
\( z = \frac{1}{{\sqrt 2} l_{B}}
(x+iy)\), the coordinate representation (using symmetric gauge 
\( {\bf A} =
\frac{1}{2}{\bf B} \times {\bf r} \) ) of the angular momentum states 
of the
lowest Landau level (LLL) is \begin{equation}
\langle z,z^{*} \mid\! l \rangle \equiv \hspace{.1cm} \langle z,z^{*} 
\mid\!
l,n=0 \rangle = \frac{1}{\sqrt{\pi l!}} z^{l} e^{-\frac{1}{2} z z^{*}} .
\end{equation} Comparing with (\ref{eq:scser}), we see that the coherent
states can be identified, up to a normalization factor, with the projected
position eigenstates of the full system,
\be
\mid \! z^{*},n=0 \rangle = \sqrt{\pi}\; P \mid \! z^{*},z \rangle ,
\;\;\;\;P=\sum_l \mid \! l\rangle \langle l \mid \! . \ee Thus, for a
general wave function of LLL, which has the form \be
\langle z,z^{*} \mid\! \psi \rangle = \psi(z) e^{-\frac{1}{2} z z^{*}}
\ee
with $\psi(z)$ as an analytic function of $z$, the variable $z$ can be
identified either as the position variable projected on LLL or,
equivalently, as the coherent state variable of the HW algebra.

When acting only on the analytic part of the wave functions, the guiding
center operators \(a,a^{\dagger}\) are in the LLL represented by
\begin{equation} a^{\dagger} = z , \hspace{.5cm} a=\frac{d}{dz},
\label{eq:lllaakors} \end{equation} and a Fock-Bargmann representation
\cite{kn:perelomov,kn:bargmann} is obtained.

As emphasized in ref.\cite{kn:girv}, restriction to the lowest Landau 
level
involves a projection for the observables with an associated operator
ordering rule. This is because although the operators \(z\) and \(z^{*}\)
commute, the same is not true for the projected operators. In fact, the
operator ordering rule is easily obtained by an algebraic approach.
According to (\ref{eq:guirel}) and (\ref{eq:comp}), the particle
coordinate operators are given by
\begin{equation} z= a^{\dagger}-b , \hspace{.5cm} z^{*}=a-b^{\dagger}.
\end{equation} Since \(b\) annihilates any state in the LLL, while
\(b^{\dagger}\) lifts the state up to the next Landau level we have \be
bP=Pb^\dagger=0
\ee with $P$ as the projection on LLL. $P$ also commutes with $a$ and
$a^\dagger$, and this implies for the projection of a product of powers of
$z$ and $z^*$, \begin{eqnarray} P z^{k}(z^{*})^{l}P &=& P
(a-b^{\dagger})^{l} (a^{\dagger}-b )^{k} P \nonumber \\ &=& a^{l}
(a^{\dagger})^{k} . \end{eqnarray} The operator ordering is unique,
and 
the
rule found in ref.\cite{kn:girv} is the coordinate representation of this
expression.

So far we have implicitly assumed that the particle is allowed to
move 
in a
plane with no boundary, in which case each Landau level is infinitely
degenerated and yields a representation of the Heisenberg-Weyl
algebra. 
For
the comparison with the quasi-hole case it may be of interest to consider
also a case where there is a boundary to the system. The simplest case
is a
circular boundary, which can be obtained by introducing a limit to the
angular momentum number, \(l \leq l_{max} = \frac{\Phi}{\Phi_{0}}\), where
\(\Phi\) is the total magnetic flux and \(\Phi_{0} =\frac{hc}{q}\) is the
flux quantum. This implies that the raising operator \(a^{\dagger}\) no
longer leaves the state space invariant. However, an operator annihilating
the uppermost state may be defined as
\begin{equation} d_{+} = \sqrt{l_{max} - L +1}\; a^{\dagger}. 
\end{equation}
With the adjoint operator \(d_{-} \equiv d_{+}^{\dagger}\) and 
\(d \equiv L
- \frac{l_{max}}{2}\) we then obtain an operator set \(\{d_{+}, d_{-}, d\}
\) spanning an $su(2)$ algebra. Thus a consequence of limiting the
area 
is that
the state space no longer defines an irreducible representation of the HW
algebra, but instead an irreducible representation of this $su(2)$ 
algebra.
The dimension of the representation is determined by the area of the 
system.
The finite value of \(l_{max}\) also implies that the coherent states must
be modified. An obvious modification is to use the spin coherent states of
the $su(2)$ algebra. Following ref.\cite{kn:perelomov} these states
are 
given
by \begin{eqnarray}
\mid\! \eta^{*} \rangle_{sc} &=&
(1+\mid\!\eta\!\mid^{2})^{-\frac{l_{max}}{2}} e^{\eta^{*}d_{+}} \mid\!
l =0
\rangle \nonumber \\ &=& (1+\mid\!\eta\!\mid^{2})^{-\frac{l_{max}}{2}}
\sum_{l =0}^{l_{ max}} (\eta^{*})^{l} \sqrt{\frac{l_{max}!}{l!(l_{
max}-l)!}} \mid\!l \rangle,
\end{eqnarray} and satisfy the equation
\begin{equation} (d_{-} + 2 \eta^{*} d - (\eta^{*} )^{2} d_{+} )\mid\!
\eta^{*} \rangle_{sc} =0.
\end{equation} Now choose \(\alpha = \sqrt{l_{max}} \eta\), and hold
\(\alpha\) fixed while the area of the system grows. Then the spin 
coherent
state \(\mid\! \eta^{*} \rangle_{sc} \) evolves into the HW-algebra 
coherent
state \( \mid\!\alpha^{*} \rangle\) as the area goes to infinity. We
also have
\begin{eqnarray}
\lim_{l_{max}\rightarrow \infty } \frac{1}{\sqrt{l_{max}}} d_{+} \mid\!l
\rangle &=& \sqrt{l +1} \mid\!l+1 \rangle_, \nonumber \\
\lim_{l_{max}\rightarrow \infty } \frac{1}{\sqrt{l_{max}}} d_{-} \mid\!l
\rangle &=& \sqrt{l} \mid\!l-1 \rangle, \nonumber \\
\lim_{l_{max}\rightarrow \infty } \frac{1}{l_{max}} d \mid\!l \rangle &=&
\mid\!l \rangle.
\end{eqnarray}
This shows explicitly how the Heisenberg-Weyl algebra is
reestablished for the infinite
system \cite{kn:perelomov}. The correspondence also implies that the 
localization property
of the HW-algebra coherent state is taken well care of by the $su(2)$ 
algebra
spin coherent state as long as the particle is not close to the 
boundary of
the system, \(\mid\!\eta\mid \ll 1\).

One should note that the coherent states defined in this way no longer are
identical to the position eigenstates projected to the restricted
area 
of the
lowest Landau level. Such a projection is obtained from the expression of
the HW coherent states (\ref{eq:scser}) by limiting the sum to $l \leq
l_{max}$.


\subsection{One quasi-hole.}

Consider \(N\) electrons in a magnetic field \({\bf B}= -B {\bf e}_{z}\)
(\(B>0\)), and assume that the system contains a single quasi-hole. The
filling factor is \(\nu =1/m\) where \(m\) is odd. The symmetric gauge is
used. Laughlin's wave function for this system is \cite{kn:laugh1}
\begin{equation}
\Psi_{m,z_{0}}(z_{1},_{\cdots}, z_{N}^{*}) =
\psi_{z_{0}}(z_{1},_{\cdots},z_{N}) \phi_{m}(z_{1},_{\cdots}, z_{N}^{*}),
\label{eq:laugh1}
\end{equation} where
\begin{equation}
\phi_{m}(z_{1},_{\cdots}, z_{N}^{*}) = e^{-\frac{1}{2}\sum_{i=1}^{N}z_{i}
z_{i}^{*}}\prod_{i<j}(z_{i}-z_{j})^{m} \label{eq:grunn} \end{equation} is
the ground state wave function and
\begin{equation}
\psi_{z_{0}}(z_{1},_{\cdots},z_{N}) = \prod_{i=1}^{N}(z_{i}-z_{0})
\label{eq:hufu}
\end{equation} is the factor which pushes the electrons away from a small
area around the point \(z_{0}\). The electron charge is by convention
\(-e\). The ground state
$\phi_m$ corresponds, for finite $N$, to an electron "droplet" of constant
density
$1/(2\pi m l_B^2)$ and of circular shape around the origin. The radius
is $l_B
\sqrt{2mN}$.

For the discussion of the one hole-states it is sufficient to consider
only
the \(\psi_{z_{0}}\)-part of the wave function. The ground state factor
$\phi_m$ does not depend on the position \(z_{0}\) of the quasi-hole 
and can
be absorbed in the integration measure. Notice that \(\psi_{z_{0}}\) does
not depend on the complex conjugate coordinates, it is an analytic 
function
of each of the \(z_{i}\)'s.

\(\psi_{z_{0}}\) will now be expanded in terms of eigenfunctions of the
angular momentum. These states define a complete set of orthogonal states
for the state space of the $N$-electron system with a single quasi-hole.
When the angular momentum operator is pulled through the ground state
\(\phi_{m}\) it picks up a term \(m\) from each pair of electrons, and
since \(\psi_{z_{0}}\) is independent of the complex conjugates the 
angular
momentum operator now is \begin{equation} L = \sum_{i=1}^{N}
z_{i}\partial_{z_{i}} + \frac{m}{2}N(N-1). \end{equation} The expansion is
done by use of the so-called {\em elementary symmetric polynomials}
\(S_{k}\). In fact,
\(\psi_{z_{0}}\) can be viewed as defining these functions as follows
\cite{kn:lang}:
\begin{equation}
\psi_{z_{0}}(z_{1},_{\cdots},z_{N}) = \prod_{i=1}^{N}(z_{i}-z_{0}) =
\sum_{k=0}^{N}(-z_{0})^{N-k}S_{k}(z_{1},_{\cdots},z_{N}) .\label{eq:eksp}
\end{equation} But this expression also is an expansion of the wave 
function
\(\psi_{z_{0}}\) in terms of angular momentum eigenstates. This is 
realized
by noting that the elementary symmetric polynomials are homogeneous of
degree
\(k\) in \(z_{1},_{\cdots},z_{N}\). Hence, 
\begin{eqnarray} L
S_{k}(z_{1},_{\cdots},z_{N}) &=& 
\left(\sum_{i=1}^{N} z_{i}\partial_{z_{i}}
+ \frac{m}{2}N(N-1) \right) S_{k}(z_{1},_{\cdots},z_{N}) \nonumber \\ &=&
\left(k+\frac{m}{2}N(N-1) \right) S_{k}(z_{1},_{\cdots},z_{N}) .
\end{eqnarray} The state space for the system of \(N\) electrons with a
single quasi-hole may then be defined as being the function space 
spanned by
the set of $N+1$ states.
\begin{equation}
\{ S_{k}(z_{1},_{\cdots},z_{N}) \}_{k=0}^{N} .
\end{equation}
(Note that $S_0$ is identical to the ground state of a system with $N$
electrons.)

The functions \(S_{k}\) are orthogonal since they all have different
eigenvalues for the total angular momentum. However, they are not
normalized. The scalar product is
\begin{equation} (S_{k},S_{k^{\prime}}) = \int d\mu (z) \left(
S_{k}(z_{1},_{\cdots},z_{N})\right)^{*}
S_{k^{\prime}}(z_{1},_{\cdots},z_{N}), \label{eq:skpro} \end{equation}
where
the integration measure is \begin{equation} d\mu (z) = d^{2}\!z_{1}\cdots
d^{2}\!z_{N}
\prod_{i<j}^{N}\!\mid \!z_{i}-z_{j}\!\mid^{2m} e^{-\sum _{i=1}^{N}\mid
z_{i}\mid ^{2}}. \label{eq:intmea}
\end{equation} For arbitrary \(m\) this scalar product indeed is 
troublesome
because of the product of all pairs, 
\(\prod \!\mid \!z_{i}-z_{j}\!\mid^{2m}
\). In fact, only for \(m=1\) we are able to calculate the norm of the
angular momentum eigenfunctions \(S_{k}\) exactly. This case 
corresponds to
fermionic quasi-holes. On the other hand, for arbitrary values of \(m\) it
is in the limit of large \(N\) possible to use the plasma analogy to find
approximate solutions for the relative normalization.

Let us briefly review how the plasma analogy \cite{kn:Lau} is used to
determine the normalization factor \(N_{z_{0}}\) for the wave function
\(\Psi_{m,z_{0}}(z_{1},_{\cdots}, z_{N}^{*})\) (\ref{eq:laugh1}). The
normalization integral then is interpreted as the classical probability
distribution for a charge (the quasi-hole) in a plasma of $N$ other 
charges
(the electrons). The integrand is written in the form
\(\mid\!N_{z_{0}} \Psi_{m,z_{0}}\!\mid^{2} \equiv e^{-\beta
U(z,z^{*},z_{0},z_{0}^{*})}\) where \(\beta\) is some constant 
(the inverse
temperature) and \(U\) is interpreted as the potential energy of the
classical system of charges.  The potential $U$ is dictated by the form
of the hole state (\ref{eq:laugh1}).  The \(N\) charges have values
\(q=\sqrt{m/\beta}\) and interact with a logarithmic Coulomb potential.
They also have a Coulomb interaction with a uniform background charge 
density
\(\sigma = -1/(\pi\sqrt{m\beta})\). The additional charge, of
value \(\bar{q} = q/m\) (the hole), has a logarithmic Coulomb interaction
with the
\(N\) other charges.  If also this
charge has a Coulomb interaction with the background charge, the
probability distribution becomes independent of the position of the charge
due to screening by the
$N$ free charges. This is interpreted as the condition for the 
normalization
integral to be independent of $z_0$. Thus, the normalization factor is
determined by the form of this interaction and is (up to a constant 
factor)
\begin{equation} N_{z_{0}} = e^{-\frac{1}{2m} \mid z_{0} \mid ^{2}}.
\label{eq:ninte} \end{equation}

The validity of this method to determine the normalization factor
rests 
upon
the assumption that the number \(N\) of electrons is a large number
and 
the
main contribution to the integral is due to the classical configuration
where the \(N\) identical particles form a droplet of uniform charge
density. One generally expects corrections when \(\bar{q}\) is near the
boundary. Now, holding both the filling fraction and the magnetic
field 
at
constant values, an increase in the number of electrons must be 
accompanied
by an increase in area. Hence, the limit \(N\rightarrow\infty\) 
effectively
means that boundary effects in the quantum Hall system are neglected.

With the normalization factor determined we can use the expansion
\[
\int d^{2}\!z_{1}\cdots d^{2}\!z_{N} \mid \! N_{z_{0}}\Psi_{m,z_{0}}
\! 
\mid
^{2} = e^{-\frac{1}{m} \mid z_{0} \mid ^{2}}
\sum_{k=0}^{N}(z_{0}z^{*}_{0})^{k} (S_{N-k}, S_{N-k}) , \] to find the
relative
norm of the angular momentum eigenfunctions. The result is
\begin{equation}
\frac{(S_{k+1},S_{k+1})}{(S_{k},S_{k})} = m \,(N-k)\hspace{1cm} (N \gg 1).
\label{eq:heidi}
\end{equation}
It is of interest to compare this result with the exact result which
can be obtained for $m=1$. The state $S_k$ then is an
$N$-electron state with all angular momentum states $l=0,1,...,N$ filled
except for a hole in the state $l=N-k$. The normalization integral then is
simply
equal to the product of the normalization integrals of the single-particle
states of the filled levels. They are of the form
\[
\int d^{2}\!z \, e^{-z z^{*}}  (z z^{*})^{l} = \pi l! . \] 
The relative norm of two
elementary symmetric polynomials (\ref{eq:skpro}) is given by the ratio
between two such integrals. In particular we have
\begin{equation}
\frac{(S_{k+1},S_{k+1})}{(S_{k},S_{k})} = (N-k) \hspace{1cm}
(m=1).\label{eq:relno1}
\end{equation}
This shows that the approximate result (\ref{eq:heidi}) derived from
the 
plasma
analogy is in fact exact for $m=1$.

Before proceeding we introduce an abstract notation for the hole wave
functions and let the normalized ket \(\mid\! k \rangle\) correspond 
to the
angular momentum wave function $S_{N-k}$. Thus, the  electron coordinate
representation of the state is (\ref{eq:grunn}, \ref{eq:eksp},
\ref{eq:skpro})
\begin{equation}
\langle z_{1},_{\cdots},z_{N}^{*} \mid\! k \rangle =
\frac{1}{(S_{N-k},S_{N-k})^{1/2}}S_{N-k} (z_{1},_{\cdots},z_{N})
\phi_{m}(z_{1},_{\cdots},z_{N}^{*}). \label{eq:absk} \end{equation}
(Note that our notation for the states  emphasizes the role of the
quasi-hole. The quantum number \(k\) of the ket \(\mid\!k\rangle\) in the
case \(m=1\) corresponds to a hole in the single-electron state
\(z^{k}e^{-\frac{1}{2}\mid z\mid^{2}}\). When we later on let
\(N\rightarrow\infty\) while holding \(k\) fixed, this means that the hole
is held at a fixed position in the plane.) Similarly, the quasi-hole wave
function (\ref{eq:laugh1}) is represented by the abstract state \(\mid \!
z_{0}\rangle _{nn}\), \begin{equation}
\langle z_{1},_{\cdots},z_{N}^{*} \mid \! z_{0}\rangle _{nn} =
\Psi_{z_{0},m}(z_{1},_{\cdots},z_{N}^{*}). \label{eq:abdef} \end{equation}
The subscript `nn' indicates that the state is not normalized, and in this
abstract notation we have
\begin{equation}
\mid \! z_{0}\rangle _{nn} = \sum _{k=0}^{N} (-z_{0})^{k}
(S_{N-k},S_{N-k})^{1/2} \mid\! k \rangle . \label{eq:abexp} \end{equation}

A unitary transformation \(U\) from the quasi-hole system onto the single
particle system can now be defined. Motivated by the physical pictures for
\(m=1\) we define for arbitrary $m$
\begin{equation} U\mid\! k\rangle_{qh} \equiv \mid\! k \rangle_{sp},
\end{equation} where we have introduced the subscript $qh$ for the hole
states and $sp$ for the single-particle states of the previous subsection.
Transforming the normalized quasi-hole state
\(\mid\!z_{0}\rangle\) we then obtain
\begin{eqnarray} U \mid\!z_{0}\rangle &=& \frac{1}{ _{nn}\langle z_{0}
\mid\! z_{0} \rangle _{nn} ^{1/2}}U \mid\!z_{0}\rangle_{nn} \nonumber
\\ 
&=&
\frac{1}{ _{nn}\langle z_{0} \mid\! z_{0} \rangle _{nn} ^{1/2}} \sum
_{k=0}^{N} (-z_{0})^{k} (S_{N-k},S_{N-k})^{1/2} \mid\! k \rangle_{sp}
\nonumber \\ &=& \left( \sum _{k=0}^{N} (z_{0}z_{0}^{*})^{k}
\frac{(S_{N-k},S_{N-k})}{(S_{N},S_{N})} \right)^{-1/2} \sum _{k=0}^{N}
(-z_{0})^{k} \frac{ (S_{N-k},S_{N-k})^{1/2} }{ (S_{N},S_{N})^{1/2} }
\mid\! k \rangle _{sp} \nonumber \\ &\approx& \left(
\sum _{k=0}^{N}
\frac{(z_{0}z_{0}^{*})^{k}}{ m^{k}k! }
\right)^{-1/2}
\sum _{k=0}^{N}
\frac{ (-\frac{z_{0}}{\sqrt{m}})^{k} }{ \sqrt{k!} } \mid \! k \rangle_{sp}
\hspace{1cm} (N\gg 1) \nonumber \\ &\rightarrow & \mid\! \alpha ^{*} =
-\frac{z_{0}}{\sqrt{m}} \rangle _{sp},  \hspace{4.5cm}  
( N\rightarrow\infty).
\label{map}
\end{eqnarray}
This shows that when boundary effects in the Hall system are neglected and
the normalization found by the plasma analogy is applied, the quasi-hole
state \(\mid\!z_{0}\rangle\) is mapped into a HW algebra coherent
state 
of the
single particle system. In this sense the system of \(N\) electrons 
with one
(quasi)hole is effectively a one-dimensional single particle system
\cite{kn:hansleinmyr}. The relation \(\alpha ^{*} = 
-\frac{z_{0}}{\sqrt{m}}\)
between the coordinates can be understood if we take into mind the 
physical
picture of the two systems. The change of sign and complex conjugation is
related to the change in sign of the charge of the hole relative to 
that of the
electrons, and the factor \(1/\sqrt{m}\), which means an
effectively larger magnetic length for the holes, determines the size
of the hole charge relative to that of the electrons.

Returning to the case of large, but finite N, we can also find the
expressions for the $su(2)$-operators of the one-hole system. They are
related
to the $su(2)$ operators of the single-particle case by the unitary
transformation $U$. It is convenient to introduce the following
raising 
and
lowering operators which leave the (N+1)-dimensional one-hole space
invariant,
\begin{equation} D_{+}^{\prime} = \sum_{i=1}^{N}\partial_{z_{i}} , 
\hspace{1cm} D_{-}^{\prime}
= \sum_{i=1}^{N}(z_{i}-z_{i}^{2}\partial_{z_{i}}) .
\label{eq:ehvse}
\end{equation}
Together with  the shifted angular momentum operator
\begin{eqnarray} D^{\prime} &=& -L+\frac{m}{2}N(N-1) +\frac{N}{2} 
\nonumber \\  &=&
-\sum_{i=1}^{N} z_{i}\partial_{z_{i}} + \frac{N}{2}, \label{eq:angmif}
\end{eqnarray}
they define an $su(2)$ algebra. However, this is a non-hermitian 
algebra. The
correctly normalized operators, which correspond to the operators
$d_{\pm}, d$ in the single-particle case are
\begin{eqnarray} D     &=& D^{\prime}, \nonumber \\ D_{-} &=&
\frac{1}{\sqrt{m}}D_{-}^{\prime}\left( \frac{N}{2} -D^{\prime}
+1\right) 
^{-1/2}, \nonumber \\
D_{+} \equiv D^{\dagger}_{-} &=& \sqrt{m}\left(
\frac{N}{2}-D^{\prime} +1 \right) ^{1/2}D_{+}^{\prime}. 
\label{eq:hsuN} \label{Dprime}
\end{eqnarray} and these define a standard representation of the $su(2)$
generators.

If the number of
electrons is increased while we consider the action of
\(D_{+},D_{-}\) and \(D\) on some state \(\mid\!
k \rangle\) with fixed value \(k\), we obtain 
\begin{eqnarray}
\lim_{N\rightarrow\infty} \frac{1}{\sqrt{N}} D_{-} \mid\! k \rangle
&=& \lim_{N\rightarrow\infty} \frac{1}{\sqrt{N}} \sqrt{k(N-k+1)} 
\mid\! k-1
\rangle \nonumber \\ &=& \sqrt{k} \mid\! k-1 \rangle , \nonumber \\
\lim_{N\rightarrow\infty} \frac{1}{\sqrt{N}} D_{+} \mid\! k \rangle
&=& \sqrt{k+1} \mid\! k+1 \rangle , \nonumber \\ \lim_{N\rightarrow\infty}
\frac{2}{N} D \mid\! k \rangle &=& - \mid\! k \rangle .
\end{eqnarray}
This shows how the HW algebra is obtained from the $su(2)$ algebra when
$N\rightarrow\infty$.

The raising and lowering operators have a rather complicated form when
expressed as electron operators. However, a hole-coordinate representation
can be defined which is identical to a coherent state representation
of the HW algebra for $N \rightarrow \infty $. For an arbitrary state
\(\mid \! \psi \rangle =\sum c_{k}\! \mid\! k\rangle\) this 
representation is
defined by \(\psi(-\frac{z_{0}^{*}}{\sqrt{m}}) \equiv e^{\frac{1}{2m}\mid
z_{0} \mid ^{2}} \langle z_{0} \mid \!\psi \rangle\). In this 
representation
the lowering operator \(A=\lim_{N\rightarrow\infty} \frac{1}{\sqrt{N}}
D_{-}\) and its hermitian conjugate gets a simple form
\begin{eqnarray}
\langle z_{0} \mid A \mid \!\psi \rangle &=& e^{-\frac{1}{2m}\mid
z_{0} 
\mid
^{2}} \frac{d}{d(-\frac{z_{0}^{*}}{\sqrt{m}})}
\psi(-\frac{z_{0}^{*}}{\sqrt{m}}) , \nonumber \\ \langle z_{0} \mid
A^{\dagger} \mid \!\psi \rangle &=& e^{-\frac{1}{2m}\mid z_{0} \mid ^{2}}
(-\frac{z_{0}^{*}}{\sqrt{m}}) \psi(-\frac{z_{0}^{*}}{\sqrt{m}}) .
\end{eqnarray} This is the Fock-Bargmann representation \cite{kn:bargmann}
for the system of \(N\) electrons with a quasi-hole.

The mapping (\ref{map}) between the quasi-hole states and the coherent
states of the single particle system was based on the (approximate) 
form of
the norm of the elementary symmetric polynomials obtained by use of the
plasma analogy. Also the form of the hermitian $su(2)$ operators
(\ref{Dprime}) was derived by use of this expression for the norm. 
However,
we can now check directly the action of these operators on the quasi-hole
states. The lowering operator of the HW algebra, when applied to the
quasi-hole state gives \be
\frac{1}{\sqrt{N}}D_{-}\sum_{k=0}^N (-z_0)^k
S_{N-k}=-\frac{z_0}{\sqrt{m}}\sum_{k=0}^{N-1} (-z_0)^k 
\sqrt{\frac{N-k}{N}}
S_{N-k}.
\ee There is a correction factor $\sqrt{\frac{N-k}{N}}$ in this 
expression,
but this tends to
$1$ for large $N$ (if the hole stays at finite distance from the 
origin when
$N$ increases). Thus, the interpretation of the quasi-hole states as
coherent states of a HW algebra is confirmed without the use of the 
detailed
form of the norm of the polynomials $S_{k}$. However, to check the
hermitian properties of the algebra the scalar product is needed, and
for this the plasma analogy has to be applied.

\newpage

\section{Two particles.}

We now turn to the case of two quasi-holes. It is useful to start by
recalling some properties of the two-anyon system, where we also include
some comments on the distinction between the anyon coordinate and the
coherent state coordinate. This distinction will be important when we 
in the
next subsection discuss the two-hole system and its relation to the anyon
system. An explicit mapping is performed.

\subsection{The two anyon system.}

Consider two anyons in a magnetic field. For a homogeneous
field the center of mass motion can be separated out, and the hamiltonian
for the relative coordinate of the two particles is identical to the
hamiltonian (\ref{eq:start}). The parameters \(m\) and \(q\) then are 
reduced
mass and charge, respectively. Denoting the anyon parameter by
\(\nu\), 
the
wave functions now should satisfy the symmetry condition
\(\psi(e^{i\pi} z) = e^{i\nu \pi}\psi(z)\), where \(z\) is the complex,
dimensionless, relative coordinate. The solution of the energy problem
then
shows that there is a (generalized) lowest Landau level where the
energy 
is
independent of \(\nu\), and the function space is spanned by the 
orthonormal
angular momentum functions
\begin{equation}
\psi_{l,\nu}(z,z^{*}) = \frac{1}{\sqrt{\pi\Gamma(2l + \nu +1)}} z^{2l+\nu}
e^{-\frac{1}{2}zz^{*}}.
\end{equation} It is convenient to restrict \(\nu\) to the interval
\(0\leq\nu <2\), which implies \(l=0,1,_{\cdots}\). Due to the anyonic
properties, there is now a shift in the eigenvalues of the (relative)
angular momentum, \begin{equation} L\psi_{l,\nu}(z,z^{*}) = (2l+\nu)
\psi_{l,\nu}(z,z^{*}). \label{eq:shift} \end{equation}

In the single-particle case the projection on the lowest Landau level 
can be
viewed as a dimensional reduction of the system. Thus, the hermitian
combinations $a+a^\dagger$ and $i(a-a^\dagger)$ are canonically conjugate
and can be viewed as the phase space variables for a particle moving 
in one
dimension. The Heisenberg-Weyl algebra generated by these operators is the
fundamental algebra of observables of the system.

For two particles in one dimension, the relative coordinate and 
momentum are
not true observables since they do not respect the symmetry of the system.
However, the quadratic combinations do, and they span the $su(1,1)$ 
algebra,
which can now be viewed as the fundamental algebra of observables. The
irreducible representations are labelled by a real parameter which is
interpreted as the one-dimensional statistics parameter. In
ref.\cite{kn:hansleinmyr} it was shown how to identify this $su(1,1)$
algebra for two anyons in the lowest Landau level. The angular momentum
operator (\ref{eq:shift}) is, up to a constant, identical to the compact
generator of the algebra, and the shift of the eigenvalues is identified
with the one-dimensional statistics parameter. In this way the algebraic
approach to the dimensional reduction can be carried out also for the
two-anyon system in the lowest Landau level. The result is a simple linear
relation between the one-dimensional statistics parameter and the
(two-dimensional) anyon parameter.

We let the set \(\{B,B_{+},B_{-}\}\) span the $su(1,1)$ algebra, with 
$B$ as
the compact generator, $B_{+}$ and $B_{-}$ as the (hermitian conjugate)
raising and lowering operators. Represented in anyon coordinates the
operators are given by ref.\cite{kn:hansleinmyr},
\begin{equation} B_{+} = \frac{1}{2} z^{2} M, \hspace{.5cm} B_{-} =
\frac{1}{2} M\frac{d^{2}}{dz^{2}} , \hspace{.5cm} B=\frac{1}{2}(L +
\frac{1}{2}), \label{eq:nesi}
\end{equation} where \(M\) is the square root of \(M^{2}\); 
\begin{equation}
M^{2} = 1- \frac{\nu(\nu -1)}{(L +1)(L +2)}, \hspace{.5cm} 
L = z\frac{d}{dz}
. \label{eq:sist}
\end{equation}
These operators are assumed to act on the non-exponential
part of the wave functions \(\psi_{l,\nu}\).
The eigenvalues of the operator $B$ in this specific representation are
given by
\be
b_k= k+\half(\nu+1)
\ee
and with the one-dimensional statistics parameter identified with the 
shift
in the spectrum, this exhibits the simple linear relation between the
one-dimensional parameter and the 2-dimensional anyon parameter $\nu$ 
referred
to above. Note however the restriction $0\leq\nu<2$ which applies to
the anyon parameter. For general hermitian representations of the
$su(1,1)$-algebra one has a less restrictive condition, $-1<\nu$. 
Values of
$\nu\geq 2$ can in fact be included also in the anyon picture, but 
only when a
repulsive interaction is introduced, which effectively excludes the
states with the lowest relative angular momenta from the space of 
states. In the
same way the values $-1< \nu<0$ can be interpreted within the anyon 
picture as
due to a short range attractive interaction which makes the wave 
functions
singular (but still normalizable) when the relative coordinate tends 
to zero.

In the case of fermions and bosons
\(M\) equals the identity, which yields \(B_{+} = \frac{1}{2}
(a^{\dagger})^{2}\), \( B_{-} = \frac{1}{2}a^{2}\) and
\(B=\frac{1}{4}(a^{\dagger}a+aa^{\dagger})\) (\ref{eq:lllaakors}).
Therefore, coherent states of the $su(1,1)$ algebra, defined as 
eigenstates of
\(B_{-}\) \cite{kn:barut}, will in the case of general \(\nu\) be
generalizations of the symmetric/antisymmetric combinations of the 
maximally
localized HW-algebra coherent states. Using an abstract bra-ket notation
where \(\mid\!l,\nu\rangle\) corresponds to the wave function
\(\psi_{l,\nu}\), these coherent states are given by \begin{equation}
\mid\!\beta,\nu\rangle_{cs} = {\cal N}_{\beta}^{(1)} \sum_{l=0}^{\infty}
\frac{ (\frac{\beta^{*}}{\sqrt 2})^{2l} }{
\sqrt{l!\Gamma(l+\nu+\frac{1}{2})}} \mid\!l,\nu\rangle \end{equation} and
satisfy
\begin{equation} B_{-} \mid\!\beta,\nu\rangle_{cs} =
\frac{1}{2}(\beta^{*})^{2} \mid\!\beta,\nu\rangle_{cs} .
\end{equation} The Berry phase corresponding to an interchange of the two
anyons has been calculated for the state \(\mid\!\beta,\nu\rangle_{cs}\),
and the asymptotic form has been shown to have a term equal to (minus) the
anyon parameter \cite{kn:hansleinmyr}. We would now like to give some
comments on questions related to coherent
states, localization and the Berry phase. The considerations will be of
importance when we later on shall discuss the quantum Hall system. In the
single-particle case there was in the lowest Landau level an 
identification
of the particle coordinate and the coherent state coordinate. In the 
present
two-particle case such an identification holds only for fermions and 
bosons.
This is readily noted by observing that a coherent state representation of
\(B_{+}\) now gives \(B_{+} = \frac{1}{2}\beta^{2}\), since \(B_{+} =
B_{-}^{\dagger}\). The anyon coordinate \(z\) rather corresponds to 
the state
\begin{equation}
\mid\!z,\nu\rangle = {\cal N}_{z}^{(2)} \sum_{l=0}^{\infty}
\frac{(z^{*})^{2 l}}{\sqrt{\pi\Gamma(2l+1+\nu)}} \mid\!l,\nu\rangle.
\end{equation} This state, which is the projection to LLL of a position
eigenvector of the full anyon system, is an eigenstate of the operator
\(M^{-1}B_{-}\);
\begin{equation} M^{-1}B_{-} \mid\!z,\nu\rangle = \frac{1}{2}(z^{*})^{2}
\mid\!z,\nu\rangle .
\end{equation} This operator is the adjoint of the operator
\(\frac{1}{2}z^{2}\) when the system is restricted to the lowest Landau
level, hence \(M^{-1}B_{-}\) is the projection onto the LLL of the 
operator
\(\frac{1}{2}(z^{*})^{2}\).

The distinction between \(\mid\!z,\nu\rangle \) and
\(\mid\!\beta,\nu\rangle_{cs}\) will not affect the asymptotic 
behaviour of
localization properties or geometric phase. However, the leading 
corrections
to the asymptotic form of the Berry phase are different. To demonstrate
this, let us first consider a general state of the form
\begin{equation}
\mid\!\eta,\eta^{*}\rangle = {\cal N}(\eta\eta^{*}) \sum_{l=0}^{\infty}
(\eta^{*})^{2l} a_{l} \mid\!l,\nu\rangle, \end{equation} where 
\(a_{l}\) are
some expansion coefficients. We parameterize \(\eta\) along a circle around
the origin, \(\eta = \eta_{0} e^{i\phi}\). The Berry connection then is
\begin{eqnarray} i\hspace{.1cm} \langle \eta,\eta^{*} \mid \partial_{\phi}
\mid\! \eta,\eta^{*} \rangle &=& - \hspace{.1cm} 
\langle \eta,\eta^{*} \mid
(\eta\partial_{\eta}-\eta^{*}\partial_{\eta^{*}}) \mid\! \eta,\eta^{*}
\rangle \nonumber \\ &=& -\eta\eta^{*} \frac{d}{d(\eta\eta^{*})} \ln
\mid\!{\cal N}(\eta\eta^{*})\mid^{2}. \label{asbe}
\end{eqnarray} We note that the normalization factor ${\cal
N}(\eta\eta^{*})$ on one hand determines the Berry phase and on the other
hand uniquely determines the expansion coefficients \(a_{l}\) and 
therefore
the state $\mid\!\eta,\eta^{*}\rangle$.

We specialize this to the case of the $su(1,1)$ algebra coherent 
states. The
normalization factor is given in terms of a modified Bessel function,
so the asymptotic behaviour is
\begin{equation}
\mid\!{\cal N}_{\beta}^{(1)}\mid^{2} = \frac{
(\frac{\beta\beta^{*}}{2})^{\nu -\frac{1}{2}} }{I_{\nu -
\frac{1}{2}}(\beta\beta^{*})} \sim \sqrt{\pi} 2^{1-\nu}
(\beta\beta^{*})^{\nu} e^{-\beta\beta^{*}} (1+\frac{\nu(\nu
-1)}{2\beta\beta^{*}}).
\end{equation} Hence,
\begin{equation} i \hspace{.1cm} \langle \beta,\nu \mid \partial_{\phi}
\mid\!\beta,\nu\rangle_{cs} \sim
\beta\beta^{*} - \nu + \frac{\nu(\nu -1)}{2 \beta\beta^{*}}
\label{eq:asycoh} \end{equation} for \( \mid\beta\mid \gg 1\). For the anyon
coordinate state the normalization factor is more complicated. 
However, it
equals a symmetrized degenerate hypergeometric function, and using 
relations
between these and the Whittaker functions \cite{kn:gradrez} we find the
asymptotic behaviour
\begin{eqnarray}
\mid \! {\cal N}_{z}^{(2)} \mid^{-2} &=& \sum _{l=0}^{\infty}
\frac{(zz^{*})^{2l}}{\pi \Gamma(2l+\nu +1)} \nonumber \\ 
\noalign{\medskip}
& \sim & \frac{1}{2\pi} \left( e^{zz^{*}} (zz^{*})^{-\nu} + 
e^{-zz^{*}} \cos
(\pi\nu)(zz^{*})^{-\nu} \right) \nonumber
\\
\noalign{\medskip} & & -\frac{\nu}{\pi \Gamma(\nu +1)} \sum_{n=0}^{\infty}
\frac{(\nu -1)(\nu -3) \cdots (\nu -2n -1)}{(zz^{*})^{2n+2}} \nonumber \\
\noalign{\medskip} &\approx & \frac{1}{2\pi} e^{zz^{*}} (zz^{*})^{-\nu}
\left( 1- \frac{2\nu(\nu -1)}{\Gamma(\nu +1)} e^{-zz^{*}} 
(zz^{*})^{\nu -2}
\right). \label{eq:noras}
\end{eqnarray} This yields
\begin{equation} i \hspace{.1cm} \langle z,\nu \mid \partial_{\phi} \mid\!
z,\nu\rangle \sim zz^{*} - \nu + \frac{2\nu(\nu -1)}{\Gamma(\nu +1)}
e^{-zz^{*}} (zz^{*})^{\nu -1}. \label{eq:asyany} \end{equation}

As advertized, the anyon coordinate state and the $su(1,1)$-coherent state
are equivalent as far as the asymptotic behaviour of the Berry phase is
concerned. The difference shows up in the leading correction to the
asymptotic form. For fermions and bosons the difference disappears, but in
the general anyon case the state \(\mid\! z,\nu \rangle\) reaches the
asymptotic form \(zz^{*} - \nu\) faster than the coherent state.

For the later discussion it is of interest to note that the specific
way in which the statistics parameter $\nu$ appears in the Berry phase
(\ref{eq:asycoh}) (and in (\ref{eq:asyany})) depends on some conventions
chosen in the anyon description. To be more precise, let us for the moment
denote the one-dimensional statistics parameter by
$\nu_1$. This is the parameter which defines the shift in the 
spectrum of the
operator
$B$ of the $su(1,1)$ algebra, and it is the same parameter as we read
from the Berry phase
(\ref{eq:asycoh}). It can furthermore be related to the asymptotic
behavior of the wave functions
$\mid\psi\mid\approx \mid z\mid^{\nu_1+2m}, m=0,1,...$, when the relative
coordinate $z$ tends to zero. (It is also the same as the exclusion
statistics parameter for anyons in the lowest Landau level
\cite{kn:excl,kn:canr,kn:wu}). With the conventions chosen here it 
is up to a
factor $1/\pi$ also identical to the phase angle obtained by the 
interchange of the two anyons,
\be
\nu_1=\frac{\theta}{\pi}\equiv \nu_2, \,\,\,\, 0\leq\theta<2\pi .
\label{stat1}
\ee
The convention implies that the orientation of the loop which 
interchanges the
two particles is fixed relative to the vector $q\vec B$, which depends on
the charge $q$ of the particles. However, the opposite orientation may be
chosen. The relation then is changed to
\be
\nu_1=2-\frac{\theta}{\pi}=2- \nu_2, \,\,\,\, 0\leq\theta<2\pi.
\label{stat2}
\ee
This implies that if the statistics angle is kept fixed, relative 
to a fixed
orientation of the interchange loop, but the sign of the charge is 
changed,
then the relation between the (two-dimensional) statistics angle and the
one-dimensional statistics parameter is changed from (\ref{stat1}) to
(\ref{stat2}).


\subsection{Two quasi-holes.}

The Laughlin wave function for two quasi-holes at positions 
\(z_{0_{1}}\) and
\(z_{0_{2}}\) is
\begin{equation}
\Psi^m_{z_{0_1},z_{0_2}}(z_{1},_{\cdots}, z_{N}^{*}) =
\prod_{i=1}^{N} (z_{i} - z_{0_{1}}) \prod_{j=1}^{N}(z_{j} - z_{0_{2}})
\phi_{m}(z_{1},_{\cdots}, z_{N}^{*}), \label{eq:laugh2} \end{equation}
where
$\phi_m$ is the Laughlin ground state (\ref{eq:grunn}). In the general
two-hole case the total
angular momentum is degenerate. However, since we shall be mainly 
concerned
with the anyonic properties of the system, which are related to the 
relative
motion of the two quasi-holes, we choose to investigate the 
restricted case
with two holes located symmetrically around the origin. We then have
$z_{0_{1}}=-z_{0_{2}}\equiv Z/2$, and the degeneracy in the total angular
momentum is removed. The hole-part of the wave function is 
\begin{eqnarray}
\psi_{Z}(z_{1},z_{2},_{\cdots},z_{N}) &\equiv& \prod_{i=1}^{N} (z_{i} -
\frac{Z}{2}) \prod_{j=1}^{N}(z_{j} + \frac{Z}{2}) \nonumber \\ &=&
\prod_{i=1}^{N} (z_{i}^{2}-(\frac{Z}{2})^{2}) \nonumber \\ &=&
\sum_{k=0}^{N} (\frac{Z}{2})^{2N-2k} (-1)^{N-k} S_{k}(z_{1}^{2},
z_{2}^{2},_{\cdots},z_{N}^{2}). \label{eq:toh} \end{eqnarray} The angular
momentum eigenstates are represented by the elementary symmetric 
polynomials
\(S_{k}\) which now are functions of the squared electron coordinates.
Defining
\(\tilde{S}_{k}(z_{1},_{\cdots},z_{N}) \equiv S_{k}(z_{1}^{2},
z_{2}^{2},_{\cdots},z_{N}^{2}) \) and using the homogeneity of 
\(S_{k}\) we
find for the angular momentum eigenvalues \begin{eqnarray}
L\tilde{S}_{k}(z_{1},z_{2},_{\cdots},z_{N}) &=& \left( \sum_{i=1}^{N}
z_{i}\partial_{z_{i}} + \frac{m}{2}N(N-1) \right) \tilde{S}_{k}(z_{1},
z_{2},_{\cdots},z_{N}) \nonumber \\ &=& \left( 2k + \frac{m}{2}N(N-1)
\right) \tilde{S}_{k}(z_{1},z_{2},_{\cdots},z_{N}) . \end{eqnarray} Since
there is no degeneracy, the angular momentum states $\tilde{S}_k$ are
orthogonal, but also in this case they are not normalized.

We introduce an abstract notation for the states of the two-hole
system. 
Let
the ket \(\mid\! k \rangle \) be defined by \begin{equation}
\langle z_{1},_{\cdots}, z_{N}^{*} \mid \!k \rangle = 
\frac{1}{\sqrt{c_{k}}}
\tilde{S}_{N-k}(z_{1},_{\cdots},z_{N}) \phi_{m} 
(z_{1},_{\cdots},z_{N}^{*}),
\end{equation} where \(\phi_{m}\) is the ground state wave function
(\ref{eq:grunn}), and let \(c_{k}\) be the norm (\ref{eq:skpro}) of the
angular momentum eigenfunction \(\tilde{S}_{N-k}\);
\begin{equation} c_{k} = (\tilde{S}_{N-k},\tilde{S}_{N-k}). 
\label{eq:antsc}
\end{equation} Similarly, let \( \mid\! Z \rangle \) be the 
normalized state
of which the coordinate representation is (apart from the normalization
factor) given by \(\psi_{Z}\phi_{m}\) (\ref{eq:grunn}, \ref{eq:toh}), 
that is
\begin{equation}
\mid\! Z \rangle = {\cal N}(ZZ^{*}) \sum_{k=0}^{N} (\frac{Z}{2})^{2k}
(-1)^{k} \sqrt{c_{k}} \mid\! k
\rangle\,
\end{equation} where
\begin{equation}
\mid\!{\cal N}(ZZ^{*})\!\mid^{-2} = \sum_{k=0}^{N} (ZZ^{*})^{2k}
\frac{c_{k}}{4^{2k}} .
\end{equation}

To determine the normalization factor \({\cal
N}(ZZ^{*})\) Laughlin's plasma analogy is again used. The
normalization 
factor
then represents the interaction potential between two (external) 
charges of
strength \( \bar q = \sqrt{\frac{1}{m\beta}}\) (the quasi-holes) 
and also the
interaction between these charges and the uniform background charge 
density
\(\sigma = -\frac{1}{\pi \sqrt{m\beta}}\). The interactions with the free
charges of the plasma are determined by the
$z_i$-dependent part of the wave function, and they have the form of
two-dimensional (logarithmic) Coulomb potentials. Therefore, if \({\cal
N}(ZZ^{*})\) is assumed to have a form determined by Coulomb interactions
between the external charges and between these charges and the background
charge, then the plasma analogy asserts that the normalization
integral 
will
be (almost) independent of \(ZZ^{*}\) due to charge screening. For small
\(ZZ^{*}\) the screening will not be complete, but this effect is expected
to vanish exponentially. There will also be a correction for values of
\(ZZ^{*}\) which brings the two charges close to the boundary of the 
plasma.
Motivated by these considerations we write the normalization factor as
\begin{equation}
\mid\!{\cal N}(ZZ^{*})\!\mid^{2} = R e^{-\frac{1}{2m}ZZ^{*}}
(ZZ^{*})^{\frac{1}{m}} (1+\Delta V(ZZ^{*})). \label{eq:tilpl} 
\end{equation}
The first two $Z$-dependent factors correspond to the Coulomb interactions
and \(\Delta V \) represents the interaction potential which survives the
screening. This potential is expected to vanish exponentially as the 
charges
are moved away from each other. $R$ is an $N$- and $m$-dependent factor
related to the ground state energy of the system.

We now return to the discussion of the localized two-anyon states. The
Berry
connections of the coherent state (\ref{eq:asycoh}) and the 
anyon coordinate
state (\ref{eq:asyany}) were shown to have the same asymptotic behaviour,
but with different corrections for finite distance between the
anyons. 
These
results we will compare with the corresponding result for the quasi-hole
system. Using (\ref{asbe}) and (\ref{eq:tilpl}), with $Z = \mid\! Z \!\mid
e^{-i\phi}$, we find for the Berry connection of the two-quasi-hole system
\begin{equation} i \langle Z \mid \partial_{\phi} \mid\!Z\rangle =
\frac{1}{2m}ZZ^{*} - \frac{1}{m} - ZZ^{*}\frac{d}{d(ZZ^{*})} \Delta
V(ZZ^{*}). \label{eq:ber}
\end{equation} The two first terms correspond to the results of
ref.\cite{kn:arovas}, and hence show that the main assumption from the
plasma analogy, namely the Coulomb interaction, is in accordance with the
independent calculation of the Berry phase. The two leading terms 
also agree
with the corresponding terms of the anyon system (both for the anyon
coordinate state and the $su(1,1)$-coherent state), provided we 
identify the
statistics parameter $\nu$ with $1/m$ and $zz^*$ with $ZZ^*/2m$, 
with $z$ as
the anyon coordinate. However, the form of the third term of 
(\ref{eq:ber}),
which is also determined by the plasma analogy, seems to agree only 
with the
result for the anyon coordinate state, but not with the coherent
state. 
The
coherent state has a correction term with a slower, non-exponential 
fall-off
with distance. In the plasma picture it would correspond to a surviving
dipole-dipole potential, which falls off like
$1/r^2$. To the best of our knowledge such a dipole interaction is not
expected to be present in the two-dimensional plasma.
For this reason the Laughlin quasi-hole state
corresponds more closely to the anyon coordinate state than to the
$su(1,1)$-coherent state.

Based on the correspondence with the anyon system we have the following
expression for the normalization factor \({\cal N}\) \begin{equation}
\mid\!{\cal N}(ZZ^{*})\!\mid^{-2} = R \sum_{k=0}^{N} \frac{
(\frac{ZZ^{*}}{2m})^{2k}}{\pi \Gamma(2k+\frac{1}{m} +1)} . \label{eq:nof}
\end{equation} It gives the correct behaviour of the equivalent plasma as
well as the expected form of the Berry phase (see (\ref{eq:noras}) and
(\ref{eq:tilpl})). Thus, for a large electron number \(N\) the expression
(\ref{eq:nof}) approximates the exact one with high accuracy when
$ZZ^{*}$ is not too small (of order the magnetic length squared). The
deviation from the exact result tends exponentially fast to zero when
$ZZ^{*}$ increases.

The formal definition of the mapping from the quasi-hole system onto the
anyon system can now be given. We introduce the subscripts $qh$ for the
quasi-hole states and $a$ for the anyon states to distinguish between the
states of this and the previous subsection. The mapping is 
\begin{equation}
U \mid\!k\rangle _{qh} = (-1)^{k} \mid\! k, \nu = \frac{1}{m} \rangle_{a}.
\end{equation}
\(U\) is a unitary operator, and we have identified the anyon parameter of
the quasi-holes as \(\nu = \frac{1}{m}\) in order to have accordance 
between
(\ref{eq:ber}) and (\ref{eq:asyany}). Then \begin{eqnarray} U \mid\! Z
\rangle_{qh} &=& {\cal N}(ZZ^{*}) \sum_{l=0}^{N} \frac{Z^{2l}}{2^{2l}}
\sqrt{c_{l}} \mid\! l, \frac{1}{m} \rangle _{a} \nonumber \\ 
&\approx& \sqrt
R {\cal N}(ZZ^{*})
\sum_{l=0}^{N} \frac{ (\frac{Z}{\sqrt{2m}})^{2l} }{\sqrt{\pi \Gamma (
2l+\frac{1}{m} +1)}} \mid\! l, \frac{1}{m} \rangle _{a} \hspace{7mm}
(N\gg 1)
\nonumber
\\ &\rightarrow& \mid\! z=-\frac{Z^{*}}{\sqrt{2m}} ,
\frac{1}{m} \rangle_{a}
\hspace{4.8cm}(N\rightarrow\infty) .
\end{eqnarray}
This gives the explicit correspondence between the two-hole system and the
anyon system with statistics parameter $\nu=1/m$. The complex 
conjugation and
the rescaling of the coordinates is related to the change in sign and
rescaling of the charge, as has already been discussed in the one-hole
case. Note however the additional factor $1/\sqrt 2$ that is due to the
reduced charge which appears in the two-anyon system.

Our mapping is based on an approximate expression for the normalization
factor \({\cal N}\), and thereby approximate expressions for the norms
\(c_{k}\) of the angular momentum eigenstates; \begin{equation} c_{k} =
\frac{R}{\pi} \frac{2^{2k}}{m^{2k} \Gamma (2k+\frac{1}{m} +1)}.
\label{eq:apck}
\end{equation} In the case \(m=1\) the latter can be calculated
exactly. 
The
result is
\begin{equation} c_{k}^{exact} = \left\{
\begin{array}{lcl} N!\pi^{N} (N+1)!! \frac{2^{2k}}{(2k+1)!} & & (0 \leq k
\leq \frac{N}{2}) , \\ N!\pi^{N} (N+1)!! \frac{2^{2k}}{(2k+1)!} - 
N!\pi^{N}
(N+1)!! \sum_{l=0}^{2k-N-1} \frac{1}{l!(2k+1-l)!}       
& & (\frac{N}{2} <
k \leq
N) ,
\end{array}
\right.
\end{equation} where we have used the notation \(
(N+1)!!=(N+1)!N!(N-1)!\cdots 1!\). In this case the approximate expression
is in fact exact for $k\leq N/2$, which means that in the limit
\(N\rightarrow\infty\) it has the correct \(k\)-dependency for all finite
values of \(k\). The $N$-dependence of the factor $R$ is also specified by
the exact result. For each specific value of \(k\) we have
investigated 
how
the \(N\) electrons occupy the single-electron states in the lowest Landau
level. We find that \(k>\frac{N}{2}\) always corresponds to having one of
the quasi-holes outside the boundary of the system, that is it is 
actually a
single-hole state. On the other hand, if \(k\leq \frac{N}{2}\) both holes
are inside the system. Hence, in the case \(m=1\) the approximative 
results
equal the correct results for all true two-hole states.

For \(m=3\) we do not have exact analytical results for 
\(c_{k}\), but numerical values
are found for up to five electrons by use of the results of
ref.\cite{kn:Myrheim}. For each specific value of \(N\) the ratio
\(\frac{c_{k}/c_{0}}{c_{k}^{exact}/c_{0}^{exact}}\) has been evaluated and
the results are listed in Table 1.
\begin{table}
\centering
\begin{tabular}{|l|l|l|l|l|} \hline\hline  & \(N=2\) & \(N=3\) & \(N=4\) &
\(N=5\) \\ \hline \(k=1\) & $ 0.42 $ & $ 0.91 $ & $ 1.08 $ & $ 1.07 $ \\
\hline \(k=2\) & $ 0.28 $ & $ 0.49 $ & $ 0.78$ & $ 1.16 $ \\ \hline 
\(k=3\)
&       & $ 0.31 $ & $ 0.52$ & $ 0.72$ \\ \hline
\(k=4\) &       &       & $ 0.34$ & $ 0.54$ \\ \hline
\(k=5\) &       &       &       & $ 0.36$ \\ \hline
\hline\hline
\end{tabular}
\caption{
\protect \footnotesize Expansion coefficients of the hole states for 
$m=3$.
The table gives a comparison between results from the plasma analogy and
numerical results ({\em exact}) for a small number of electrons. The
entries are $(c_k/c_0)/(c_k^{exact}/c_0^{exact})$, and there is a tendency
to obtain
the correct \(k\)-dependency as \(N\) grows. The exact results are taken
from ref.\protect\cite{kn:Myrheim}.} \label{tab}
\end{table} Although the number of electrons in these examples is small,
there seems to be a tendency to reach the correct \(k\)-dependency for
$c_{k}$, \ie the value $1$ for the ratio tabulated, as \(N\) grows.

For the finite system we can, similarly to the case with one
quasi-hole, find ladder operators leaving the space spanned by \(
\{\tilde{S}_{k}(z_{1},z_{2},_{\cdots},z_{N}) \}_{k=0}^{N}\)
invariant. These are given by
\begin{eqnarray}
&C_{-} = \sum_{i=1}^{N} ( z_{i}^{2} - \frac{1}{2} z_{i}^{4}
\partial_{z_{i}}^{2}), \hspace{1cm} C_{+} = \sum_{i=1}^{N} \frac{1}{2}
\partial_{z_{i}}^{2},& \nonumber \\
\noalign{\medskip}
& C = -\frac{1}{2}
\sum_{i=1}^{N} z_{i}\partial_{z_{i}} + \frac{N}{2}, & \label{C-op}
\end{eqnarray}
and span a non-hermitian $su(2)$ algebra.

The unitary mapping between the (\(N\rightarrow\infty\)) two-hole state
space and the state space for two anyons in the lowest Landau level,
determines the form of the generators of the $su(1,1)$ algebra in the
quasi-hole case. Expressed in terms of the operators $C_\pm$ and
$K\equiv C+N/2$ they have the form
\begin{eqnarray}
\tilde{B}_{-} &=& \frac{1}{m} C_- (K-N -1)^{-1} \sqrt{\frac{K(K +
\frac{1}{m}
-\frac{1}{2})}{(K +\frac{1}{2m})(K +\frac{1}{2m} -\frac{1}{2})}} ,
\nonumber \\
\noalign{\medskip}
\tilde{B}_{+} &=& -mC_{+}(K+1)^{-1} \sqrt{(
K+\frac{1}{2m}+1)(K+\frac{1}{2m}+\frac{1}{2} )(K+1)(K+\frac{1}{m}
+\frac{1}{2})} .
\end{eqnarray} The expressions fully demonstrates that the electron
coordinate representation of the ladder operators $\tilde{B}_{+}$ and
$\tilde{B}_{-}$, which are mutually adjoint in the limit
\(N\rightarrow\infty\), is rather complicated. However, as already
discussed, the quasi-hole state \(\mid\!Z\rangle\) is not identified as a
coherent state of this algebra, but rather as an eigenvector of the 
operator
\(U^{\dagger}M^{-1}B_{-}U\) (\ref{eq:nesi}, \ref{eq:sist}). For this 
operator we have the slightly simpler
form \begin{eqnarray} U^{\dagger}M^{-1}B_{-}U &=& \frac{1}{m} C_{-} (K-N
-1)^{-1} \nonumber \\ &=& -\frac{2}{m} \sum_{i=1}^{N} (z_{i}^{2} -
\frac{1}{2} z_{i}^{4} \partial_{z_{i}}^{2}) \left( \sum_{i=1}^{N}
z_{i}\partial_{z_{i}} +2 \right)^{-1} .
\end{eqnarray}

\newpage


\section{Quasi-electrons.}

In the case of Laughlin quasi-electrons an anyon representation is not
easily obtained. This was noted also in ref.\cite{kn:laugh2}. We will 
in this
section present some of the difficulties and make some additional comments
to the problem.

We start out by considering a single quasi-electron, for which the
variational wave function
\begin{equation}
\langle z_{1},_{\cdots}, z_{N}^{*} \mid \! z_0^- \rangle =
\Psi_{m,z_0}^-(z_{1},_{\cdots}, z_{N}^{*})= \tilde{\cal{N}}(\mid\! z_0
\!\mid) e^{-\frac{1}{2}\sum_{i=1}^{N}z_{i} z_{i}^{*}} \prod_{i=1}^{N}
(\partial_{z_{i}} - z_{o}^{*}) \prod_{i<j}(z_{i}-z_{j})^{m} 
\label{eq:kvapa}
\end{equation} has been proposed \cite{kn:laugh1}. The normalization
integral for this state is
\begin{equation} 1 \equiv \langle z_0^- \! \mid \! z_0^- \rangle = \int
\prod_{i=1}^{N} d^{2}\!z_{i} \, e^{-\sum_{j=1}^{N}z_{j} z_{j}^{*}}
\prod_{k<l}\mid\! z_{k} - z_{l}\! \mid ^{2m} \prod_{i=1}^{N} ( \mid\! z_{i}
- z_{0}\! \mid^{2} -1)\mid\!
\tilde{\cal{N}}(\mid\! z_0\! \mid)\!\mid^2. \label{eq:noikvp} 
\end{equation}
In the language of the plasma analogy, there are now corrections to 
the pure
(logarithmic) Coulomb interaction between the external charge (the
quasi-electron) and the free charges (the electrons). If these corrections
are neglected one simply has the normalization integral of the single
quasi-hole, and $\tilde{\cal{N}}(\mid\! z_0 \!\mid)$ equals
(\ref{eq:ninte}). Using (\ref{asbe}) (which still is valid since
(\ref{eq:kvapa}) can be expanded in terms of orthogonal angular momentum
states), with the parameterization $z_0=\mid\!z_0\!\mid e^{-i\phi}$, 
one then
obtains a quasi-electron charge $q_{qe}=-e/m$. This
equals the result proposed in ref.\cite{kn:arovas}. In
the large $N$ limit it is then possible to map the quasi-electron state
$\mid\!z_0^-\rangle$ onto a (HW algebra) coherent state of an ordinary
single-particle system, the coherent state coordinate now being
$\alpha^*=z_0^*/\sqrt{m}$ (see (\ref{map}) for comparison).

There is however an objection to this approximation: For the full Landau
level ($m=1$) it gives the wrong expression both for the normalization
factor and for the Berry phase. Since $\Psi_{m,z_0}^-$ (\ref{eq:kvapa}) is
entirely in the lowest Landau level, it describes a situation with 
no excess
charge anywhere for $m=1$. The wave function is simply that of the ground
state, and the Berry phase is zero.

We will now show that the approximation mentioned above is equivalent to
neglecting some of the terms in the Berry connection. The latter can, when
assuming the normalization factor depends only on $\mid\!z_0\!\mid$, be
written
\be
\frac{d\gamma^-}{d\phi} & \equiv & i\langle z_0^-\mid
\frac{\partial}{\partial \phi}
\mid z_0^- \rangle \nonumber \\ &=& i \mid\!\tilde{\cal{N}}(\mid\! z_0
\!\mid)\!\mid^2 \int \prod_{i=1}^{N} d^{2}\! z_{i} \,
e^{-\sum_{i=1}^{N}z_{i}
z_{i}^{*}} \left( \prod_{j=1}^{N} (\partial_{z_{j}^*} - z_{o})
\prod_{k<l}(z_{k}^*-z_{l}^*)^{m} \right. \nonumber \\ & & \hspace{2cm}
\left.\sum_{i=1}^{N} (-\frac{dz_0^*}{d\phi}) 
(\partial_{z_{i}} - z_{o}^{*})^{-1}
\prod_{j=1}^{N} (\partial_{z_{j}} - z_{o}^{*})
\prod_{k<l}(z_{k}-z_{l})^{m}\right) . \label{eq:berf}
\ee 
The operator $(\partial_{z} - z_{o}^{*})^{-1}$ is not well defined in
single particle spaces of infinite dimension. However, we may define the
operator
\begin{equation} A \equiv -
\frac{1}{z_0^*}\sum_{l=0}^{\infty}\frac{1}{(z_0^*)^l}
\frac{\partial^l}{\partial
z^l}, \label{eq:defA}
\end{equation} 
which for any finite value of $k$ satisfies \begin{equation}
A(\partial_{z} - z_{o}^{*}) z^k = (\partial_{z} - z_{o}^{*}) A z^k = z^k.
\end{equation} Laughlin's quasi-electron wave function contains only 
finite
polynomials, hence $A$ may be used to find the Berry connection. The
details of the calculation are shown in an appendix, and the following
expression appears: 
\be
\frac{d\gamma^-}{d\phi} &=& i\int d^2\!z \rho_{z_0}^-(z,z^*)
\frac{dz_0^*}{d\phi}\frac{1}{z_0^*-z^*} \nonumber \\ &+& i\pi
\frac{1}{z_0^*}\frac{dz_0^*}{d\phi}\sum_{l=0}^L \frac{1}{(z_0^*)^l}\left(
(\partial_z +z^*)^l \rho_{z_0}^-(z,z^*) \right)_{z=z_0,z^*=z_0^*},
\label{eq:bepha} 
\ee 
where $\rho_{z_0}^-(z,z^*)$ is the single particle
density $N \int d^2z_2 \cdots d^2z_N \mid\!\Psi_{m,z_0}^-\!\mid^2$. The
upper limit of the summation index is $L\equiv m(N-1)$ and 
corresponds to the
highest power in the polynomials arising. The expression
(\ref{eq:bepha}) is valid for any
filling fraction $1/m$. The first term corresponds to the
results of ref.\cite{kn:arovas}. Assuming $\rho_{z_0}^-$ equals $1/m\pi$,
which is the ground state density for $N$ large, this term gives the Berry
phase $-2\pi \mid z_0\!\mid^2/m$, which in turn gives the charge and
normalization factor mentioned above. The second term gives an 
additional contribution to the Berry phase, and hence corrections to the
normalization factor. We conclude (for arbitrary $m$) that to neglect 
this
term is equivalent to use the approximation $\mid\! z_{i} - z_{0}\!
\mid^{2} -1 \rightarrow \mid\!z_{i} - z_{0}\! \mid^{2}$ in
(\ref{eq:noikvp}).

Using the exact expression for $\rho_{z_0}^-$ in the case
$m=1$, we have calculated the sum in (\ref{eq:bepha}). Taking the 
large $N$ limit we then obtain $2\pi\! \mid\!
z_0\!\mid^2
-2\pi N$ for this contribution to the Berry phase. Hence, we obtain the
correct phase and also the correct $\mid\!z_0\!\mid$-dependence of the
normalization factor\footnote{There is a subtlety here. 
Eq.(\ref{eq:bepha})
rests upon the assumption $\tilde{\cal{N}}=\tilde{\cal{N}}(\mid\! z_0
\!\mid)$. For $m=1$ this implies a nonphysical singularity
$\frac{(-z_0^*)^N}{\mid z_0 \mid^N}$ in the wave function. This 
singularity
is responsible for the artificial $-2\pi N$ in the Berry phase. Such
singularities are not present for other filling fractions.}, but at 
the same
time we have explicitly demonstrated that the additional term does 
not vanish
when $N\rightarrow\infty$. One might argue that $m=1$ is a special
case, but we see no obvious reason why the corrections can be 
overlooked for
other values of $m$, while they are important for $m=1$. This implies some
uncertainty in the derivation of the charge of the quasi-electrons 
from the
Berry phase.

Proceeding to the case of two quasi-electrons at $z_a$ and $z_b$, the
normalization integral is written on the same form as 
(\ref{eq:noikvp}) but
with the substitution \begin{equation}
\mid\! z_{i} - z_{0}\! \mid^{2} -1 \hspace{2mm} \rightarrow \hspace{2mm}
\mid\! z_{i} -z_{a}\!\mid ^{2} \mid\! z_{i} -z_{b}\!\mid ^{2} -4 \! \mid\!
z_{i} -\frac{1}{2}(z_{a}+z_{b}) \!\mid ^{2} +2 . \end{equation}
Using the plasma analogy language, there are again corrections to the
pure Coulomb interaction, which are also reflected in the Berry 
connection. With $z_a = \mid\!z_a\!\mid
e^{-i\phi}$, $z_b$ fixed, the latter is given by (\ref{eq:bepha}) when
substituting
\begin{equation}
\rho_{z_o}^-(z,z^*) \hspace{2mm} \rightarrow \hspace{2mm}
\rho_{z_a,z_b}^-(z,z^*),
\end{equation}
$\rho_{z_a,z_b}^-$ being the single particle density when two
quasi-electrons are present. We write
$\rho_{z_a,z_b}^- = \rho_{z_a}^- + \delta\rho_{z_b}^-$, where
$\delta\rho_{z_b}^- = \rho_{z_b}^- -\rho_0 $ is the deviation from the
ground state density due to the quasi-electron at $z_b$. For sufficiently
large separation between the quasi-electrons we assume that
$\delta\rho_{z_b}^-(z,z^*)$ as well as all its derivatives  are 
vanishingly
small at the point $z=z_a$, and derive the following approximate 
expression for the Berry phase,
\be
\gamma^- &=& i\int_0^{2\pi}d\phi \int d^2\!z \, \rho_{z_a}^-(z,z^*)
\frac{dz_a^*}{d\phi}\frac{1}{z_a^*-z^*} \nonumber \\ &+& i\pi
\int_0^{2\pi}d\phi \, \frac{1}{z_a^*}\frac{dz_a^*}{d\phi}\sum_{l=0}^L
\frac{1}{(z_a^*)^l}\left( (\partial_z +z^*)^l \rho_{z_a}^-(z,z^*)
\right)_{z=z_a,z^*=z_a^*} \nonumber \\ &-& 2\pi \int_{\mid z \mid < 
\mid z_a
\mid} d^2\! z \, \delta\rho_{z_b}^-, \label{eq:likn}
\ee where Cauchy's residue theorem has been used on the last term.  We
notice that no new types of correction terms
arise as compared to those of (\ref{eq:bepha}). The two
first terms determine the Aharonov-Bohm charge $q_{qe}$ of the
quasi-electrons and give no information about interchange phases. We
assume, as usual, this charge to be equal to the charge determined by
$\delta\rho_{z_{b}}^-$. 
This implies that the statistics phase is
\be
\nu_{qe}=\frac{q_{qe}}{e},
\label{nu-qe}
\ee
where $-e$ is the electron charge, and we note that this is the 
same relation between charge and statistics
as for the quasi-holes \cite{kn:arovas}. When the complications 
referred to
above in the derivation of the quasi-electron charge is neglected 
this gives
$\nu_{qe}=-\frac{1}{m}$.

One should note that the negative value of $\nu_{qe}$, as discussed 
in section
3.1, implies that if the quasi-electrons are interpreted as anyons in a
magnetic field, they are not free anyons, but anyons with a special
short-range attraction which effectively gives a negative statistics
parameter.

It is of interest to compare the result (\ref{nu-qe}) with numerical
simulations which have been performed for a small number of
interacting electrons in a
strong magnetic field on a sphere \cite{kn:canright}. By state 
counting the one-dimensional
statistics parameter (exclusion statistics parameter), for $m=3$, 
has there
been determined to be
$\nu_1 = 2-\frac{1}{m}$. The fractional part agrees with the value 
cited above,
but not the integer part. (To be certain that this part is correctly
accounted for one has to check that the Berry phase is well behaved 
also for
$z\rightarrow0$). A possible explanation for this is that the correction
terms we have neglected are important for the (integer part of the)
statistics parameter. However, numerical calculation of the Berry phase
indicates that the result (\ref{nu-qe}) may in fact be correct for the
Laughlin quasi-electron state {\cite{janm}. Therefore, a more likely
explanation is that the Laughlin state does not reproduce correctly this
aspect of the physical quasi-electron state of the quantum Hall system.

\section{Conclusion.}

In summary, we have examined the anyon representation of Laughlin's
quasi-particle states with emphasis on one-dimensional, algebraic aspects.
The motivation has been to give a detailed account of the correspondence
to the one-dimensional description of anyons in the lowest Landau
level.

In particular, we have formulated explicit mappings from the state
spaces of one and two quasi-holes to the state spaces of a single
particle and two anyons in the lowest Landau level. We have further
examined the question whether the Laughlin states can be viewed as
coherent states of the fundamental algebra of observables for these
systems. In the case of a single hole the Laughlin hole-state was
found to correspond to a coherent state for the one-dimensional
Heisenberg-Weyl algebra in the limit where the electron number $N$
tends to infinity. In the case of two quasi-holes we showed that
the two-hole state corresponds more closely to the anyon position 
eigenstate
projected onto the lowest Landau level than to the coherent state of the
underlying $su(1,1)$ algebra. The two-dimensional parameter was 
identified as
$1/m$ and the charge of the anyons as $-1/m$ times the electron charge.

However, one should note that the difference between the projected anyon
position eigenstate and the coherent state only shows up in the
corrections to the asymptotic form of the Berry phase for particle
interchange, and the corrections tend to zero for large separation. The
physical relevance of these correction terms is not so clear, and in
particular it is not clear whether the Laughlin hole state gives a good
representation of the physical hole state of the interacting electron
system in this respect.

For both the one-hole and the two-hole systems we have considered the 
(edge)
effects of a finite electron number $N$. The state spaces then have finite
dimensions and define hermitian representations of the compact $su(2)$
algebra rather than of the non-compact Heisenberg-Weyl or $su(1,1)$ 
algebra
which are relevant for the infinite dimensional cases. The unitary mapping
between the hole system and  the single particle/anyon system then
determines the form of the generators. The operators have a quite
complicated form when written in electron coordinates. However, in the
hole-representation the operator expressions take a simple form.

We have compared the approximative results (found from plasma analogy
and Berry phase calculations) for the norm of the angular momentum
eigenstates to known, exact results. For $m=1$ analytic expressions can be
found and these show exact agreement for all single-hole states as well as
for all true two-hole states. In the case of two quasi-holes we have 
made a
comparison to  numerical results previously obtained for
$m=3$ and a small number of electrons. We find a reasonably good agreement
when this number is increased.

For quasi-electrons explicit single-particle and two-anyon 
representations are
not so easily obtained. We have pointed out that to perform a specific
approximation in the plasma analogy calculation is equivalent to
neglect an additional contribution to the Berry phase. We have also 
shown that this extra 
term is
important for the filling fraction $m=1$. Whether it is important
for other values of $m$ is unsettled.

\vspace{15mm}

\centerline{\large\bf Acknowledgements.}
\vspace{.5cm}

We would like to thank G.S. Canright, J. Myrheim and D. Arovas for
several useful discussions.


\appendix
\section{Berry phase calculation for quasi-electrons.}

In this appendix we show the detailed derivation of the Berry
connection (\ref{eq:bepha}) for Laughlin quasi-electrons. We start 
from (\ref{eq:berf})
and (\ref{eq:defA}),
\be
\frac{d\gamma^-}{d\phi} &=& i  N \frac{1}{z_0^*} \frac{dz_0^*}{d\phi} 
\int \prod_{i=1}^{N} d^{2}\!z_{i} \mid\!\tilde{\cal{N}}(\mid\! z_0
\!\mid)\!\mid^2 e^{-\sum_{i=1}^{N}z_{i}
z_{i}^{*}} 
\left( \prod_{j=1}^{N} (\partial_{z_{j}^*} - z_{o})
\prod_{k<n}(z_{k}^*-z_{n}^*)^{m} \right.
 \nonumber \\ & & \hspace{2cm}
\left.\sum_{l=0}^{\infty} \frac{1}{(z_0^*)^l}\partial_{z_1}^l
\prod_{j=1}^{N}
(\partial_{z_{j}} - z_{o}^{*})
\prod_{k<n}(z_{k}-z_{n})^{m} \right).
\ee 
In the expression $\prod_{j=1}^{N} (\partial_{z_{j}} - z_{o}^{*})
\prod_{k<n}(z_{k}-z_{n})^{m}$ the highest power of any single electron
coordinate is $m(N-1)\equiv L$. Thus the infinite sum over $l$ is cut
off, and integrating by parts we obtain
\be
\frac{d\gamma^-}{d\phi} &=& i  N \frac{1}{z_0^*} \frac{dz_0^*}{d\phi}
\int d^2\!z_1 \sum_{l=0}^{L} (\frac{z_1^*}{z_0^*})^l \int
\prod_{i=2}^{N}d^2\!z_i \mid\!\Psi_{m,z_0}^- \!\mid^2 \nonumber \\
&=& i   \frac{1}{z_0^*} \frac{dz_0^*}{d\phi} \int d^2\!z \rho_{z_0}^-
(z,z^*) \sum_{l=0}^{L} (\frac{z^*}{z_0^*})^l \nonumber \\
&=& i \int d^2\!z \rho_{z_0}^-(z,z^*) \frac{dz_0^*}{d\phi}
\frac{1}{z_0^* -z^*} \nonumber \\
&+& i \frac{1}{(z_0^*)^{L+1}} \frac{dz_0^*}{d\phi} 
\int d^2\!z \rho_{z_0}^-
(z,z^*) \frac{(z^*)^{L+1}}{z^* -z^*_0} .
\ee 
Here $\rho_{z_0}^-(z_1,z_1^*) =N \int\prod_{i=2}^N d^2\!z_i
\mid\!\Psi_{m,z_0}^- \!\mid^2$ is the single electron density. The
first integral corresponds to the result of ref.\cite{kn:arovas}, but
we have also obtained an additional contribution to the Berry
connection. To further investigate this extra term we use the notation
\newpage
\be
p(z_1,z_1^*) &\equiv & (z_1^*)^{L+1} \int \prod_{i=2}^{N}d^2\!z_i
 \, e^{-\sum_{i=2}^{N}z_{i}z_i^*} \left( \prod_{j=1}^{N} 
(\partial_{z_{j}^*} - z_{o})
\prod_{k<n}(z_{k}^*-z_{n}^*)^{m}\right.  \nonumber \\
& & \hspace{17mm} \left.\prod_{j=2}^{N} (\partial_{z_{j}} -
z_{o}^{*}) \prod_{k<n}(z_{k}-z_{n})^{m}\right),
\ee
which is a polynomial in $z_1$ and $z_1^*$. It is important to 
realize
that the highest power of $z_1$ is $m(N-1) =L$ whereas the lowest
power of $z_1^*$ is $L+1$. Using the expression for the single
electron density we find
\be
\int d^2\!z \, \rho_{z_0}^- (z,z^*) \frac{(z^*)^{L+1}}{z^* -z^*_0} 
&=& N
\mid\!\tilde{\cal{N}} \!\mid ^2 \int d^2\!z \,
e^{-z z^*} \frac{1}{z^*
-z_0^*} (\partial_z - z_0^*) p(z,z^*) \nonumber \\
&=& N \mid\!\tilde{\cal{N}} \!\mid ^2 \int d^2\!z \,
\partial_z \left(
e^{-z z^*} \frac{1}{z^* -z_0^*} p(z,z^*) \right) \nonumber \\
&+& N \mid\!\tilde{\cal{N}} \!\mid ^2 \int d^2\!z \,
e^{-z z^*} p(z,z^*).
\ee
Integration by parts has been used to arrive at the last equality. The
last integral vanishes because there exist no matching
powers of $z$ and $z^*$ in the polynomial $p(z,z^*)$. However, the surface
integral is nonvanishing due to the pole at the position of the 
quasi-electron. We remove a small area $A$ around $z_0$. Integration over
the remaining part of the plane is performed by using Stokes theorem
to rewrite the integral to a line integral over the boundary of
$A$. We then find
\begin{equation}
\lim_{A \rightarrow 0} \int d^2\!z  \, \partial_z  \left( e^{-z z^*}
\frac{1}{z^* -z_0^*} p(z,z^*) \right) = -\pi e^{-z_0 z_0^*}
p(z_0,z_0^*).
\end{equation}
The polynomial $p(z,z^*)$ is related to the single electron density by
\begin{equation}
N \mid\!\tilde{\cal{N}} \!\mid ^2 p(z,z^*) = - (z^*)^{L+1} e^{z z^*}
\frac{1}{z_0^*} \sum_{l=0}^L \frac{1}{(z^*)^l} ( \partial_z + z^*)^l
\rho_{z_0}^- (z,z^*),
\end{equation}
so the final result for the Berry connection then is
\be
\frac{d\gamma^-}{d\phi} &=& i\int d^2\!z \, \rho_{z_0}^-(z,z^*)
\frac{dz_0^*}{d\phi}\frac{1}{z_0^*-z^*} \nonumber \\ &+& i\pi
\frac{1}{z_0^*}\frac{dz_0^*}{d\phi}\sum_{l=0}^L \frac{1}{(z_0^*)^l}\left(
(\partial_z +z^*)^l \rho_{z_0}^-(z,z^*) \right)_{z=z_0,z^*=z_0^*}.
\ee


\end{document}